\documentclass[intlimits,twoside,a4paper]{article}     % Us  CMP

\usepackage{amssymb, amsmath}

\usepackage{graphicx}          % -- Usenko
\usepackage[T2A]{fontenc}       % Us  CMP
\usepackage[cp1251]{inputenc}   % Us  CMP

\usepackage{cmpj2}

\issue{2016}{19}{3}{33603}
\doinumber{10.5488/CMP.19.33603}

\title[Adsorption-induced surface normal relaxation]%
{Adsorption-induced surface normal relaxation of a solid adsorbent}

\author{A.S.~Usenko}
\address{Bogolyubov Institute for Theoretical Physics of the National Academy of
Sciences of Ukraine, 14~b~Metrolohichna St., 03680  Kyiv,
Ukraine}

\newcommand{\sign}{\mathop{\rm sign}\nolimits}       % LATEX  2.09
\newcommand{\erfc}{\mathop{\rm erfc}\nolimits}       % LATEX  2.09
\authorcopyright{A.S.~Usenko, 2016}
\date{Received January 25, 2016}

\begin{document}

\maketitle

\begin{abstract}
We investigate adsorption of a gas on the flat surface of a solid
deformable adsorbent taking into account thermal fluctuations and
analyze in detail the effect of thermal fluctuations on the
adsorbent deformation in adsorption. The condition for coexistence
of two states of a bistable system of adsorbed particles is
derived.  We establish the specific properties of the
adsorption-induced surface normal relaxation of an adsorbent
caused by thermal fluctuations. The mean transition times between
two stable states of the bistable system are derived in the
parabolic approximation and in the general case.
\pacs   68.43.-h,  68.43.Mn,  68.43.Nr,  68.35.Rh   % Up to six PACS numbers
\end{abstract}

%%-------------------------------- Section 1 ----------------------------------

\section{Introduction}  \label{Introduction}

Investigation of adsorption of gas particles on the surfaces of solids is very
important for solving various problems of physics and chemistry. Even a monolayer
coverage of the adsorbent surface by adsorbed particles (adparticles) is capable of
considerably changing adsorbent properties (see, e.g.,
\cite{Mor,Nau78,JaP,KiK,Zhd,LNP,AdC,Nau03}).

It is also important to know the amount of adsorbed substance on the surfaces of various adsorbents
and its distribution over the adsorbent surface for
heterogeneous catalysis \cite{Bar,Rog,Roz,Bor,KSh}.  %\looseness=-1

Various generalizations of the classical Langmuir model of adsorption provide new
specific features for the amount of adsorbed substance and its kinetics (see,
e.g.,
\cite{Zhd,Rog,Roz,Bor,KSh,Rut,Tov,RuE,Do,KSt}).
For example, taking account of lateral interactions between adparticles leads
to a hysteresis loop of adsorption isotherms and to different structural changes
of the adsorbent surface
\cite{Zhd,LNP,AdC,Nau03,Rut,Tov,RuE,Do,KSt,IEr,Imb1,Imb2,Ert}. %\looseness=-1

Hysteresis-shaped adsorption isotherms also occur due to the
adsorbent deformation in adsorption. The effect of
adsorption-induced deformation of porous solids is well known
[see, e.g., \cite{Tva} (chapter~4 and references therein)].
Specific nonmonotonous behavior of the adsorption-induced
deformation of various porous adsorbents with the gas pressure is
observed in a series of experiments; theoretical explanation of
this phenomenon and its effect on the adsorption isotherms are
given in \cite{RN06,GN10,GN11}. For a solid ideal (nonporous)
adsorbent with the flat energetically homogeneous surface, a
hysteresis of adsorption isotherms of an adsorbed one-component
gas caused by the adsorbent deformation in adsorption is
established in \cite{Us12}. It should be noted that the hysteresis
of adsorption isotherms due to retardation effects in the case
where the typical adsorption--desorption time is much less than
the relaxation time of the adsorbent surface was
predicted by Zeldovich \cite{Zeld} as early as in 1938. %\looseness=-2

Memory effects are also essential in the study of the surface diffusion of
adparticles over the adsorbent surface if the typical time of the moving
adparticles is less than the relaxation time of the adsorbent surface (see,
e.g., \cite{AFY} and references therein). %\looseness=-1

For a bistable system of adparticles on the flat surface of a deformable
adsorbent, it is of great interest to investigate possible transitions between
stable states of a system. One of the ways of correct description of
these transitions is taking account of fluctuations in a system. This problem
is closely connected with investigation of the effect of fluctuations on the
normal displacement of the adsorbent surface caused by adsorption, especially,
taking into account the experimentally established phenomenon of the
adsorption-induced surface normal relaxation of an adsorbent (see, e.g., the
review \cite{Nau78} and references therein).  Only a few experimental data
for the adsorption-induced normal displacement of the flat surface of an
adsorbent are available for some specific values of the surface coverage.
According to \cite{Us12}, the theoretical dependence of this displacement on
the dimensionless gas concentration can be both a continuous function and a
discontinuous function depending on the value of the coupling parameter. The
second case corresponds to the bistable system under study. %\looseness=-1

In the present paper, we generalize the model of adsorption of a
gas on the flat surface of a deformable solid adsorbent proposed
in \cite{Us12} for the case taking into account thermal
fluctuations in the system. In section~\ref{Model}, we present
general relations for the normal displacement of the adsorbent
surface in adsorption and the amount of adsorbed substance on the
surface of an adsorbent whose desorption properties vary due to
the adsorbent deformation and analyze in detail the important case
of the system with two energetically equivalent states. In
section~\ref{Stationary}, we study the effect of thermal
fluctuations on specific features of the probability density of
the position of the adsorbent surface for monostable and bistable
systems. The mean transition times of the bistable system between
its stable states are investigated in the parabolic approximation.
In the general case, the corresponding times are obtained in
Appendix. %\looseness=-1

%\newpage

%%-------------------------------- Section 2 ----------------------------------

\section{Statement of the problem and general relations}  \label{Model}

We consider localized monolayer adsorption of a one-component gas on the flat
surface of a solid adsorbent within the framework of the model taking into
account the adsorbent deformation in adsorption \cite{Us12}. Gas particles are
adsorbed on adsorption sites with identical adsorption activities located at
the adsorbent surface. Furthermore, each adsorption site can be bound with one
gas particle. The total number of adsorption sites $N$ is constant. The
Cartesian coordinate system with the $Ox$-axis directed into the adsorbent
perpendicularly to its surface is introduced so that the gas environment and
the adsorbent with clean surface occupy the regions $x < 0$ and
$x \geqslant 0$, respectively. %\looseness=-2

First we briefly describe the model and the main results in
\cite{Us12} that are necessary
in what follows. %\looseness=-1

Each adsorption site is simulated by a one-dimensional linear oscillator that
oscillates perpendicularly to the adsorbent surface about its equilibrium
position ($x = 0$ in the absence of an adsorbate). Owing to the binding of a gas
particle with a vacant adsorption site, the spatial distribution of the charge
density of the adsorption site changes. Furthermore, this change depends on
different factors connected with the adsorbent and gas particles (see, e.g.,
\cite{Nau78,LNP,IEr,Imb1,Imb2,Ert,Nau94}). %\looseness=-2

As a result, the interaction of the bound adsorption site (the adsorption site
occupied by an adparticle) with the neighboring atoms of the adsorbent on the
surface and in the subsurface region changes, thus changing the resulting
force of the neighboring atoms acting on the bound adsorption site. This can be
described in terms of an adsorption-induced force
$\vec{F}^\textrm{a}(\vec{r},t)$, where $\vec{r}$ is the running coordinate of
the adsorption site that acts on the bound adsorption site. Due to this force,
the equilibrium position of the adsorption site ($x = 0$ for a vacant
adsorption site) shifts. Once the adparticle leaves the adsorption site, under
some conditions, another gas particle can occupy the adsorption site before it
relaxes to its nonperturbed equilibrium position $x = 0$. Thus, a gas particle
is adsorbed on the adsorbent surface with changed desorption characteristics
caused by a local deformation of the adsorbent by the previous adparticle,
which can be interpreted as adsorption with memory effect. Assume that the
adsorption-induced force $\vec{F}^\textrm{a}(\vec{r},t)$ is normal to the
boundary and depends only on the coordinate  $x$: \
$\vec{F}^\textrm{a}(\vec{r},t) \equiv \vec{F}^\textrm{a}(x,t) = \vec{e}_x\,
F^\textrm{a}(x,t)$, where $\vec{e}_x$ is the unit vector along the $Ox$-axis.
%\looseness=-2

In the model, the time-step force $\vec{F}^\textrm{a}(x,t)$ acting on the
adsorption site during discrete time intervals, where the site is occupied by an
adparticle, is replaced by an effective time-continuous adsorption-induced force
$\vec{F}^{\textrm{eff}}(x,t) = \vec{F}^\textrm{a}(x)\, \theta(t)$ taking into
account the presence of an adparticle on the adsorption site in the mean. Here,
$\vec{F}^\textrm{a}(x) = \vec{e}_x\,F^\textrm{a}(x)$ is the adsorption-induced
force acting on the adsorption site permanently bound to an adparticle and
$\theta(t) = N_\textrm{b}(t)/N$ is the surface coverage by the adsorbate and
$N_\textrm{b}(t)$ is the number of bound adsorption sites at the time $t$.
Expanding $F^\textrm{a}(x)$ in the Taylor series in the neighborhood of $x = 0$
and keeping only the first term of the expansion, in terms of the potential,
$F^\textrm{a}(x) = - \rd V^\textrm{a}(x)/\rd x$, we have %\looseness=-1
\begin{equation}
 V^\textrm{a}(x) \approx -\chi\, x\,,
\end{equation}
\noindent  where
 \[
\left. \chi = - \frac{\rd V^\textrm{a}(x)}{\rd x} \right|_{x = 0}
 \]
is the constant adsorption-induced force acting on the bound adsorption site.
%\looseness=-1

In this mean-field approximation, the kinetics of the surface coverage $\theta$
and the normal displacement~$x$ of the plane of adsorption sites, which
coincides with the coordinate of a bound adsorption site, in localized
adsorption, is described by the system of nonlinear differential equations
%\looseness=-2
\begin{eqnarray}
\label{2}
 && \left\{
\begin{array}{l}
  \alpha \dfrac{\rd x}{\rd t} + \varkappa \,x = \chi\, \theta,
      \vspace{3mm}  \\
 \dfrac{\rd \theta}{\rd t} = k_\textrm{a} C\, (1-\theta)  - k_\textrm{d}(x)\, \theta.
\end{array}\right.
\end{eqnarray}
\noindent  Here, $\alpha$ is the friction coefficient, $\varkappa$ is the
restoring force constant, $C$ is the concentration of gas particles that is
kept constant,
\begin{equation}
 k_\textrm{a} = k_+ \,\exp{\left(-\frac{E_\textrm{a}}{k_\textrm{B} T} \right)}\,, \qquad
 k_\textrm{d}(x) = k_-\, \exp{\left[-\frac{E_\textrm{d}(x)}{k_\textrm{B} T} \right]}\,
\end{equation}
are the rate coefficients for adsorption of gas particles and desorption of
adparticles, respectively, $k_+$ and $k_-$  are the preexponential factors,
$E_\textrm{a}$ and $E_\textrm{d}(x) = E_\textrm{d} + \chi\, x\,$ are the
activation energies for adsorption and desorption, respectively, $E_\textrm{d}$
is the activation energy for desorption of adparticles from the surface of a
nondeformable adsorbent ($\chi = 0$), $T$ is the absolute temperature, and
$k_\textrm{B}$ is the Boltzmann constant. %\looseness=-2

The activation energy for desorption $E_\textrm{d}(x)$ depends on the
coordinate $x$ due to the adsorbent deformation in adsorption caused by the
displacement of the equilibrium positions of bound adsorption sites from $x =
0$. The quantity $k_\textrm{d}(x)$ can be rewritten in the form %\looseness=-2
\begin{equation}
 k_\textrm{d}(x) = k_\textrm{d}\, \exp{\left(-\frac{\chi\,x }{k_\textrm{B} T} \right)},
 \label{4}
\end{equation}
\noindent where the first factor on the right-hand side of
(\ref{4})
\begin{equation}
 k_\textrm{d} = k_-\, \exp{\left(-\frac{E_\textrm{d}}{k_\textrm{B} T} \right)}
\end{equation}
\noindent is the classical rate constant for desorption of adparticles in the
Langmuir case, which is independent of the gas concentration $C$, and the
second factor shows a variation in the desorption characteristic of the adsorbent in
adsorption of gas particles on its surface.

In the general case, the activation energy for desorption also depends on the
surface coverage $\theta$ due to lateral interactions between adparticles (see,
e.g., \cite{Zhd,AdC,Tov,Do}). However, even in the absence of lateral
interactions between adparticles, the model used shows the essential difference
of the amount of the adsorbed substance on the deformable adsorbent from the
classical Langmuir results. %\looseness=-2

The first equation of system (\ref{2}) describes the motion of the
bound adsorption site in the overdamped approximation ignoring the
inertial term of the equation of motion of the oscillator of mass
defined both by the mass of adsorption site $m_0$ and by the mass
of adparticle $m.$ This approximation is true if
\cite{AVKh} %\looseness=-2
\begin{equation}
 \tau^2_M  \ll \tau^2_{\text{r}}\,,
\end{equation}
\noindent where  $\tau_M = \sqrt{M/\varkappa}$, $M = m_0 +m$, and
$\tau_{\text{r}} = \alpha/\varkappa$ is the relaxation time of an
overdamped oscillator.

The second equation of system (\ref{2}) is the classical Langmuir
equation for the kinetics of the surface coverage generalized to
the case taking into account
the adsorbent deformation in adsorption.  %\looseness=-2

In terms of the dimensionless coordinate of oscillator (or, which
is the same, the dimensionless normal displacement of the plane of
adsorption sites) $\xi = x/x_{\textrm{max}}\,$, where
$x_{\textrm{max}} = \chi/ \varkappa$ is the maximum stationary
displacement of the oscillator for the total surface coverage
($\theta = 1$) from its nonperturbed equilibrium position $x = 0$,
system (\ref{2})
takes the form  %\looseness=-2
\begin{eqnarray}
 && \left\{
\begin{array}{l}
  \alpha \dfrac{\rd \xi}{\rd t} = \varkappa \, \bigl( \theta - \xi
  \bigr),
      \vspace{3mm}  \\
 \dfrac{\rd \theta}{\rd t} = k_\textrm{a} C\, (1-\theta) - k_\textrm{d}\,\theta\, \exp{\left(-g\,\xi\,
 \right)}\,,
\end{array}\right.
\label{7}
\end{eqnarray}
\noindent  where the dimensionless quantity
\begin{equation}
g = \frac{|V^\textrm{a}|}{k_\textrm{B} T}\,, \label{8}
\end{equation}
which is called a coupling parameter of adparticles with adsorbent caused by
the adsorption-induced deformation of the adsorbent or, briefly, a coupling
parameter, has the physical meaning of the maximum increment of the activation
energy for desorption of adparticles (normalized by $k_\textrm{B} T$) due to
the adsorbent deformation in adsorption, $\ V^\textrm{a} \equiv
V^\textrm{a}(x_{\textrm{max}})= -\chi^2/\varkappa$. %\looseness=-2

Below, we use the model for the case where the variables $\xi(t)$
and $\theta(t)$ are slow and fast, respectively, i.e.,
$\tau_{\text{r}} \gg \tau_{\theta}\,$, where $\tau_{\text{r}}$ and
$\tau_{\theta} = \tau_\textrm{a}
\tau_\textrm{d}/(\tau_\textrm{a}+\tau_\textrm{d})$ are,
respectively, the relaxation times of the coordinate $\xi(t)$ and
the surface coverage $\theta(t)$ in the linear case, $
\tau_\textrm{d} = 1/k_\textrm{d}$ is the classical Langmuir
residence time of adparticles (or the typical lifetime of a bound
adsorption site) and $ \tau_\textrm{a} = 1/(k_\textrm{a} C)$ can
be regarded as the typical lifetime of a vacant adsorption site.
Following the principle of adiabatic elimination \cite{Hak} of the
fast variable $\theta(t)$
in (\ref{7}), we have %\looseness=-2
\begin{equation}
 \theta = \frac{\ell}{\ell + \exp(-g\, \xi)}\,,
 \label{9}
\end{equation}
and the coordinate  $\xi(t)$ is a solution of the differential equation
\begin{equation}
 \alpha \frac{\rd\xi}{\rd t} = F(\xi)
 \label{10}
\end{equation}
\noindent  that describes the motion of an overdamped oscillator under the
action of the nonlinear force
\begin{equation}
 F(\xi) = -\frac{\rd U(\xi)}{\rd\xi} = \varkappa \, \left[-\xi +
  \frac{\ell}{\ell + \exp(-g\, \xi)} \right],
 \end{equation}
\noindent  where  $\ell = C\, K$ is the dimensionless concentration of gas
particles and  $K = k_\textrm{a}/k_\textrm{d}$ is the adsorption equilibrium
constant in the linear case ($\chi = 0$).

The potential $U(\xi)$ can be represented in the form
\begin{equation}
 U(\xi) = \frac{\varkappa}{2} \, V(\xi), \qquad\quad
 V(\xi) = \xi^2 - 2\,\xi - \frac{2}{g}\, \ln{\frac{\ell +
 \exp{\left(-g\, \xi\, \right)}}{\ell + 1}}\,.
 \label{12}
\end{equation}
Relations (\ref{9}) and (\ref{10}) correctly describe the behavior
of $\xi(t)$ and
$\theta(t)$ for times $t \gg \tau_{\theta}\,$. %\looseness=-1

Sometimes, instead of equation~(\ref{10}), it is more convenient
to use the equation of motion of an overdamped oscillator in terms
of the coordinate $x$,
\begin{equation}
 \alpha \frac{\rd  x}{\rd t} = F(x),
 \label{13}
\end{equation}
\noindent  where the force $F(x)= -\dfrac{\rd U(x)}{\rd x}$  and
its potential $U(x)$ are expressed in terms of $F(\xi)$  and
$U(\xi)$ as follows~\cite{Chr}:
\begin{eqnarray}
 F(x)  &=&  x_{\textrm{max}}\, F(\xi)\big|_{\,\xi = x/x_{\textrm{max}}}
   = -\varkappa \, x + \chi\, \frac{\ell}{\ell + \exp(-b\, x)}\,,
  \\ [\smallskipamount]
 U(x)  &=&  x_{\textrm{max}}^2\, U(\xi)\big|_{\,\xi = x/x_{\textrm{max}}}
   = \frac{\varkappa\, x^2}{2} - \chi\, x  - k_\textrm{B} T \, \ln{\frac{\ell +
 \exp{\left(-b\, x \right)}}{\ell + 1}}\,,
 \label{15}
\end{eqnarray}
\noindent  where $b = \chi/(k_\textrm{B} T) =
g/{x_{\textrm{max}}}\,$.

In the stationary case, the first equation of system (\ref{7})
yields
\begin{equation}
\theta = \xi\,, \label{16}
\end{equation}
and the equilibrium position of the oscillator $\xi$ describing the stationary
displacement of the adsorbent surface in adsorption is a solution of the
equation
\begin{equation}
 \ell = \frac{\xi}{1 - \xi}\, \exp{\left( - g\, \xi \right)}\,.
 \label{17}
\end{equation}

According to \cite{Us12}, the behavior of system (\ref{7})
essentially depends on the values of the control parameters $\ell$
and $g$. For $g \leqslant g_\textrm{c} =4$, the system is
monostable (the single-valued correspondence between the
concentration $\ell$ and the coordinate $\xi$ occurs). For $g >
g_\textrm{c}\,$, the system is monostable only for $\ell \notin
[\ell^\textrm{b}_1, \, \ell^\textrm{b}_2]$, where
$\ell^\textrm{b}_1$ and $\ell^\textrm{b}_2$ are the
bifurcation concentrations defined by the relations %\looseness=-2
\begin{equation}
 \ell^\textrm{b}_k = \left( g \, \xi^\textrm{b}_k - 1 \right)\, \exp{\left(-g \, \xi^\textrm{b}_k\right)},
  \qquad  k = 1,2,
  \label{18}
\end{equation}
\noindent  where
\begin{equation}
 \xi^\textrm{b}_1 = \frac{1 + d}{2}\,, \qquad
 \xi^\textrm{b}_2 = \frac{1 - d}{2}
\end{equation}
\noindent  are the bifurcation values of  $\xi$, which are two-fold stationary
solutions of system (\ref{7}), the quantity  %\looseness=-1
\begin{equation}
 d = \sqrt{1 - \frac{1}{q}}
\end{equation}
\noindent is the width of the interval of instability of the system symmetric
about $\xi = 1/2$, and  $q = g/4$.  %\looseness=-1

For any  $\ell \in (\ell^\textrm{b}_1, \, \ell^\textrm{b}_2)$
called the interval of bistability, equation~(\ref{17}) has three
real solutions $\xi_1 < \xi_2 < \xi_3\,$; furthermore, the
stationary solutions $\xi_1$ and $\xi_3$ of system (\ref{7}) are
asymptotically stable (i.e., stable nodes) while the stationary
solution
$\xi_2$ is unstable (i.e., saddle). %\looseness=-1

\begin{figure}[!b]     % Us
 \centering{
 \includegraphics[width=0.4\textwidth]{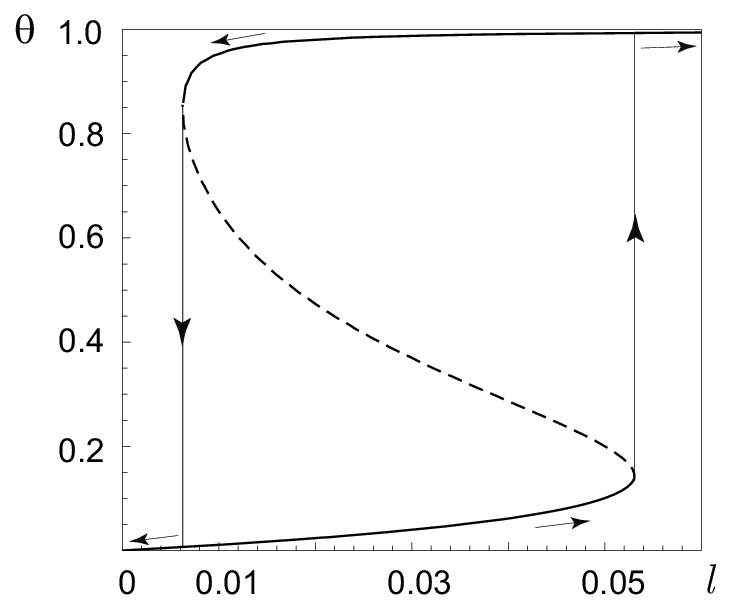}%width=130mm, height=110mm]{pfig_1}
 \caption {\label{pfig1} Adsorption isotherm for  $g$ = 8.}
 }
\end{figure}

\begin{figure}[!t]
 \centering{
 \includegraphics[width=0.4\textwidth]{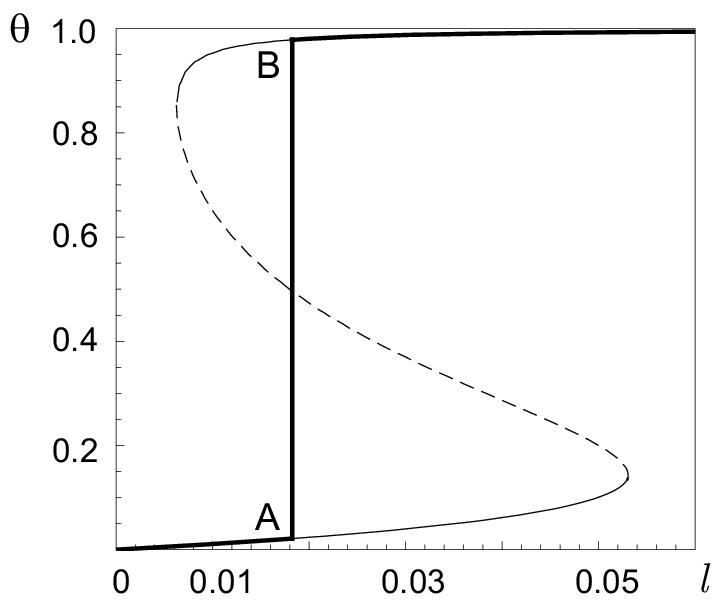}
 \caption {\label{pfig5} Adsorption isotherm (thin curve) and the Maxwell adsorption isotherm
  (thick curve) for  $g$ = 8.}
 }
\end{figure}

For $g > 4$, the adsorption isotherm $\theta(\ell)$ has a
hysteresis \cite{Us12}. An example of such an adsorption isotherm
is shown in figure~\ref{pfig1}. In this figure and in
figure~\ref{pfig5}, parts of curves of the surface coverage
$\theta(\ell)$ corresponding to stable and unstable stationary
solutions are shown by solid and broken lines, respectively. In
view of (\ref{16}), the curves in figures~\ref{pfig1} and
\ref{pfig5} also describe the displacement $\xi(\ell)$ of the
adsorbent surface from its nonperturbed equilibrium position $\xi
= 0$ with concentration $\ell$, i.e., the adsorption-induced
surface normal relaxation of a solid
adsorbent. %\looseness=-1

The surface coverage $\theta$ increases with the concentration
$\ell$ along the lower stable branch of the isotherm ending at the
bifurcation concentration $\ell = \ell^\textrm{b}_2$ (here,
$\ell^\textrm{b}_1 \approx 0.0063$ and $\ell^\textrm{b}_2 \approx
0.053$); furthermore, the increment of $\theta$ depends both on an
increase in the gas concentration and on a variation in desorption
properties of the adsorbent surface caused by the adsorbent
deformation. The jump of $\theta$ to the upper stable branch of
the isotherm at $\ell = \ell^\textrm{b}_2$ is caused exclusively
by a change in desorption properties of the adsorbent. For
convenience, transitions between stable branches of $\theta(\ell)$
are shown in figure~\ref{pfig1} by light vertical straight lines
with arrows indicating the direction of transition. Arrows under
and above stable branches of $\theta(\ell)$ show the direction of
variation in $\ell$. For $\ell > \ell^\textrm{b}_2\,$,  the
surface coverage $\theta$ varies
with $\ell$ along the upper stable branch. %\looseness=-2

The transition of $\theta$  from the lower stable branch to the upper one at
the bifurcation concentration  $\ell^\textrm{b}_2$ is accompanied by a step
increase in the activation energy for desorption, which hampers the desorption of
adparticles from the surface. As a result, as the concentration $\ell$
decreases from a value greater than $\ell^\textrm{b}_2$, the reverse transition
of $\theta$ from the upper stable branch to the lower one occurs at the lower
bifurcation concentration $\ell^\textrm{b}_1 < \ell^\textrm{b}_2$\, at which
the upper stable branch ends. %\looseness=-2

This behavior of the surface coverage $\theta$ versus the gas concentration
$\ell$ corresponds to the well-known principle of perfect delay \cite{PoS,Gil}.

The essentially different behavior of adsorption isotherms for $g
\leqslant 4$ and $g
> 4$ depends on the varying shape of the function $V(\xi)$ (also called
a potential) in these cases, namely \cite{Us12}:  $V(\xi)$ has a
single well for $g \leqslant 4$ and for $g > 4$, $\ell \notin
[\ell^\textrm{b}_1, \, \ell^\textrm{b}_2]$ and two wells with
local minima at $\xi = \xi_1$ (the first well) and $\xi = \xi_3$
(the second well) separated by a maximum at $\xi =\xi_2\,$, where
$\xi_k\,$, $k = 1,2,3$, are the solutions of equation~(\ref{17})
for $g
> 4$, $\ell \in (\ell^\textrm{b}_1, \, \ell^\textrm{b}_2)$. In the last case, denote
$V_k \equiv V(\xi_k)$, where $k = 1,2,3$, and, for a given $g$,
the position of the absolute minimum of the potential $V(\xi)$
depends on the value of $\ell$. In the special case where system~(\ref{7})
has a two-fold stationary solution ($\xi^\textrm{b}_1$
for $\ell = \ell^\textrm{b}_1$ or $\xi^\textrm{b}_2$ for $\ell =
\ell^\textrm{b}_2$), the potential $V(\xi)$ has a point of
inflection at $\xi = \xi^\textrm{b}_k\,$, $k = 1,2$, lying to the
right (for $k = 1$) or to the left (for $k = 2$) of the bottom of
the single well of  $V(\xi)$.
%\looseness=-2

We now dwell in detail on the bistable system with two energetically equivalent
states ($V_1 = V_3$). For any $g > 4$, this important case occurs for the
concentration $\ell_\textrm{M}$ defined as follows: %\looseness=-2
\begin{equation}
\ell_\textrm{M} = \exp\Bigl(-\frac{g}{2}\Bigr)\,. \label{21}
\end{equation}
\noindent Following
\cite{PoS,Gil}, the quantity $\ell_\textrm{M}$ for $g > 4$ may be
called a Maxwell concentration. It is directly verified that the potential
$V(\xi)$ with $\ell = \ell_\textrm{M}$ denoted by $V^\textrm{M}(\xi)$,
\begin{equation}
 V^\textrm{M}(\xi) = \widetilde{\xi}^{\,2} - \frac{1}{4} - \frac{1}{2\,q}\,
  \ln{\dfrac{\cosh{2\,q\,\widetilde{\xi}}}{\cosh{q}}}\,,  \qquad
  \widetilde{\xi} = \xi - \frac{1}{2}\,,
  \label{22}
\end{equation}
\noindent  is an even function about $\xi = 1/2$. Hence, the
function $V^\textrm{M}(\xi)$ (called a Maxwell potential) at $\xi
= 1/2$ has either the minimum (a single well for $q \leqslant 1$)
or the maximum (two wells for $q > 1$)
%\looseness=-1
\begin{equation}
 \widetilde{V}^\textrm{M} \equiv  V^\textrm{M}\Bigl(\xi = \frac{1}{2}\Bigr) =
 \frac{\ln{\cosh{q}}}{2\,q} - \frac{1}{4}\,.
 \label{23}
\end{equation}
The double-well Maxwell potential $V^\textrm{M}(\xi)$ has equal
minimal values denoted by $V^\textrm{M}_{\textrm{min}}$  at $\xi =
\xi_1 \equiv \xi^\textrm{M}_-$ and $\xi = \xi_3 \equiv
\xi^\textrm{M}_+\,$, where $\ \xi^\textrm{M}_\pm = (1 \pm
\eta/q)/2\,$ and $\eta$ is a positive solution of the equation
\begin{equation}
 \frac{x}{q} = \tanh{x}\,.
 \label{24}
\end{equation}
This equation directly follows from (\ref{17}) with  $\ell =
\ell_\textrm{M}$  or from the condition that the force  $F(\xi)$
with  $\ell = \ell_\textrm{M}$ denoted by $F^\textrm{M}(\xi)$,
\begin{equation}
 F^\textrm{M}(\xi) = \frac{\varkappa}{2}\, \Bigl (
 \tanh 2\,q\,\widetilde{\xi} - 2\,\widetilde{\xi} \Bigr )\,,
  \qquad   \widetilde{\xi} = \xi - \frac{1}{2}\,,
\end{equation}
is equal to zero. According to relations (\ref{22}) and
(\ref{23}), the wells are separated by a barrier of the height
$\,\Delta_\textrm{M} = \widetilde{V}^\textrm{M} -
V^\textrm{M}_{\textrm{min}}\,$ relative to the bottoms of the
wells,
%\looseness=-1
\begin{equation}
\Delta_\textrm{M} = \frac{1}{2\, q}\, \Bigl (\ln{\cosh{\eta}} - \frac{\eta^2}{2\,q}\,
\Bigr )\,.
\end{equation}
In the case where the coupling parameter $g$  is close to the
critical $g_\textrm{c}\,$, i.e., $q = 1 + \varepsilon, \ 0 <
\varepsilon \ll 1$, we obtain $\eta \approx \sqrt{3\,
\varepsilon}$ and the very low barrier $\Delta_\textrm{M} \approx
3\, \varepsilon^2/8$ separating the closely spaced
wells. %\looseness=-1

For $q \gg 1$, we get  $\eta \approx q\, (1 - 2\,
\ell_\textrm{M})$, which yields $\xi^\textrm{M}_+ \approx 1 -
\ell_\textrm{M}\,$, $\xi^\textrm{M}_- \approx \ell_\textrm{M}$ and
\begin{equation}
 V^\textrm{M}_{\textrm{min}} \approx -\ell_\textrm{M}^{\, 2}\,,  \qquad
 \widetilde{V}^\textrm{M} \approx \frac{1}{4} - \frac{\ln 2}{2\,q}\,.
\end{equation}
In this case, the wells are far spaced and the barrier height
$\Delta_\textrm{M} \approx \widetilde{V}^\textrm{M} $  tends to its maximum
value equal to 1/4 as $g\rightarrow \infty $. %\looseness=-1

\begin{figure}[!t]
 \centering{
 \includegraphics[width=0.4\textwidth]{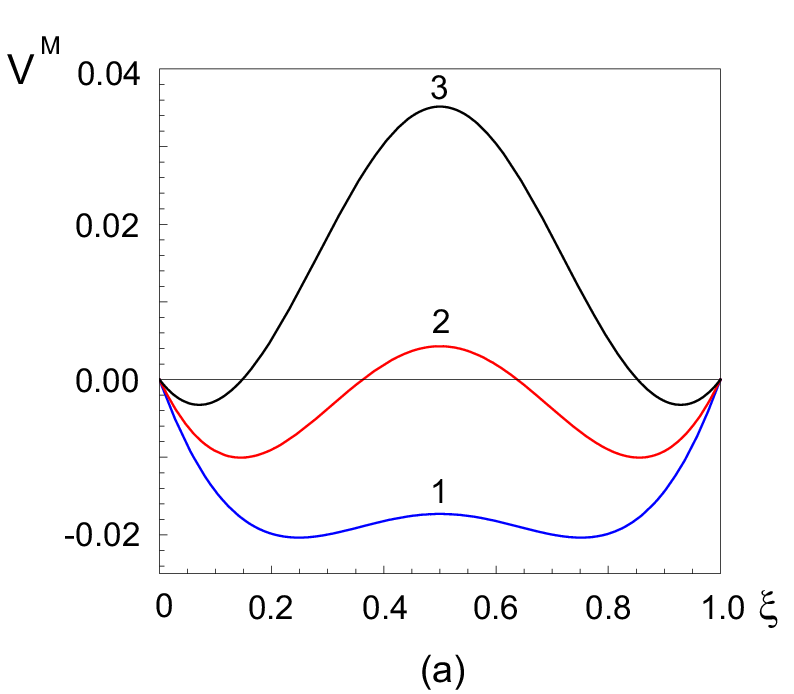}%[width=60mm, height=90mm]{pfigc_2a}
 \hspace{0.5cm}
 \includegraphics[width=0.4\textwidth]{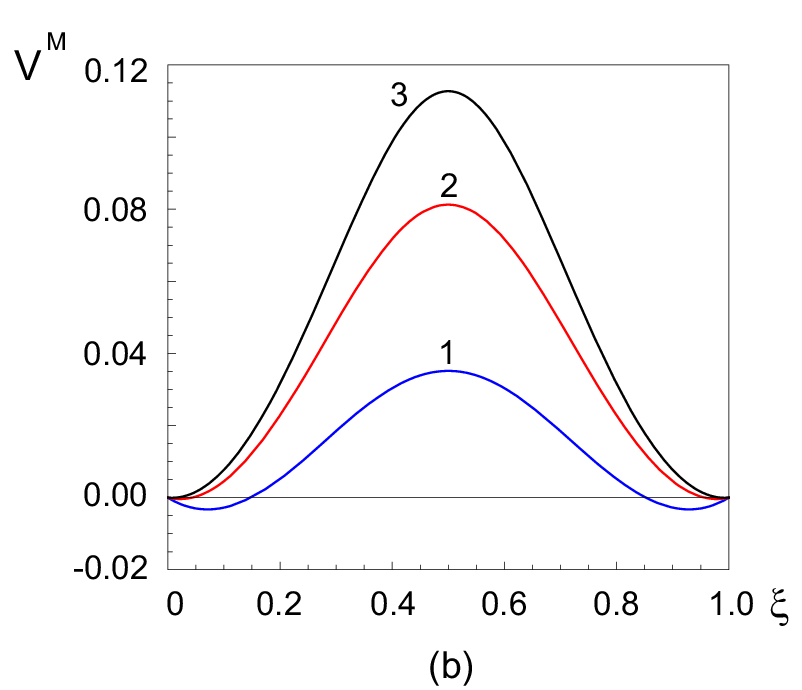}%[width=60mm, height=90mm]{pfigc_2b}  \\
 \caption {\label{pfig2} (Color online) The Maxwell potential  $V^\textrm{M}(\xi)$ for:
  (a)\, $g = $  4.4~(1),  5~(2),  6~(3); \
  (b)\, $g = $  6~(1),    8~(2), 10~(3).}
 }
\end{figure}

\begin{figure}[!h]
 \centering{
 \includegraphics[width=0.4\textwidth]{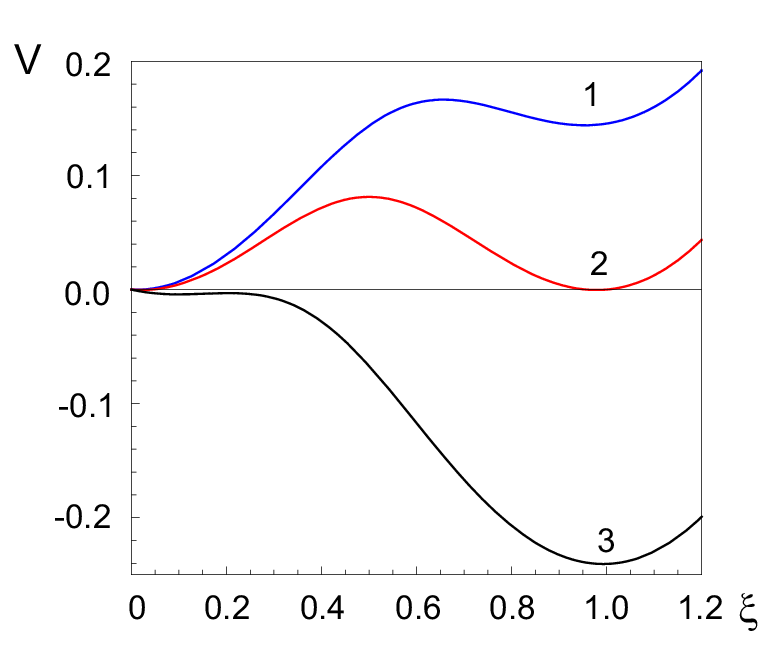}
 \caption {\label{pfig3} (Color online) The potential  $V(\xi)$ for $g = 8$  and the concentration
 $\ell = $ 0.01~(1), $\ell_\textrm{M}$~(2), 0.05~(3).}
 }
\end{figure}

The curves in figure~\ref{pfig2} show the behavior of the
Maxwell potential $V^\textrm{M}(\xi)$ with the coupling parameter~$g$, namely: the growth of the barrier between the
wells and the motion of the wells from $\xi = 1/2$.  %\looseness=-2

The curves in figure~\ref{pfig3} depicted for $g = 8$ and $\ell
\in (\ell^\textrm{b}_1, \, \ell^\textrm{b}_ 2)$, where
$\ell^\textrm{b}_1 \approx 0.0063$ and $\ell^\textrm{b}_2 \approx
0.053$, clearly illustrate transformations of the double-well
potential $V(\xi)$ with $\ell$. For $\ell \in (\ell^\textrm{b}_1,
\, \ell_ \textrm{M})$, where $\ell_\textrm{M} \approx 0.0183 $,
the first well is deeper than the second one (curve 1). Thus, the
system is stable in the first well and metastable in the second
one.  As $\ell \in (\ell^\textrm{b}_1, \, \ell_\textrm{M})$
increases, the depths of both wells increase with $\ell$ but with
different increments. For $\ell = \ell_\textrm{M}\,$, the
double-well potential is symmetric about $\xi = 1/2$ (curve 2).
For $\ell \in (\ell_\textrm{M}\,, \, \ell^\textrm{b}_2)$, the
second well is deeper than the first (curve 3). Hence, the system
is metastable in the first well and stable in the second one.
However, according to the principle of perfect delay
\cite{PoS,Gil}, the oscillator, which was initially at rest in the
first well, remains in this well with an increase in $\ell$ until
the well disappears. For the transition of the oscillator from the
first well into the deeper second well following the Maxwell
principle \cite{PoS,Gil}, it is necessary to take into account
additional factors that enable the oscillator to overcome the
barrier between the wells, e.g., fluctuations or the inertia
effect (as in \cite{Us09Pr}). In the next section of the paper, to
investigate transitions of the oscillator from one well into
another, we take thermal
fluctuations in the system into account. %\looseness=-2

\begin{figure}[!t]
 \centering{
 \includegraphics[width=0.4\textwidth]{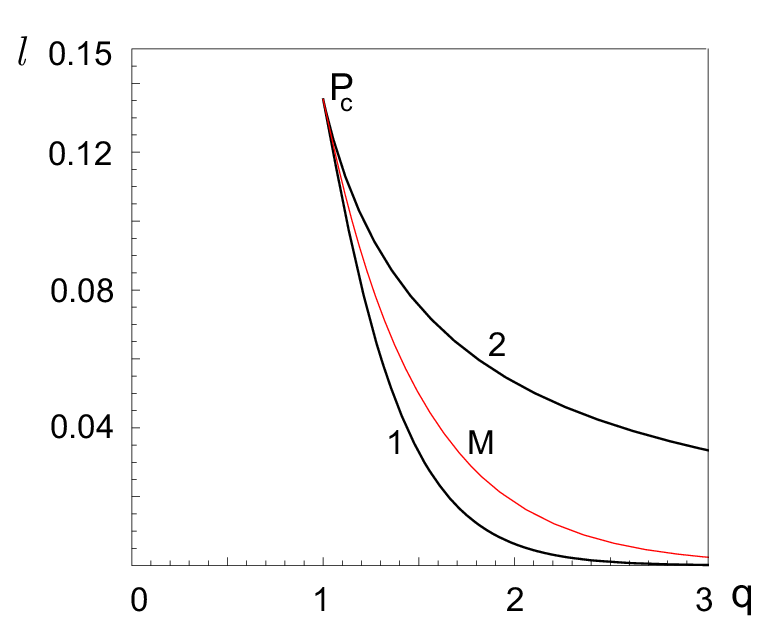}
 \caption {\label{pfig4} (Color online) Bifurcation curve (branches 1 and 2 of the curve correspond to the
  bifurcation concentrations $\ell^\textrm{b}_1$  and $\ell^\textrm{b}_2\,$, respectively) and
  the Maxwell set (curve M).}
 }
\end{figure}

The bifurcation curve in the plane of control parameters $(q,
\ell)$ in figure~\ref{pfig4} divides the first quadrant of the
plane into two parts. Branches 1 and 2 of the curve correspond to
the bifurcation concentrations~$\ell^\textrm{b}_1$ and
$\ell^\textrm{b}_2\,$, respectively, defined by relations
(\ref{18}) and curve M is the Maxwell set of the values of the
concentration $\ell_\textrm{M}$ for $q > 1$ defined by (\ref{21}).
The domain outside the bifurcation curve is a domain of
monostability of the system; for any point of this domain, system
(\ref{7}) has one asymptotically stable stationary solution. The
open domain enclosed by the bifurcation curve is a domain of
bistability of the system; for any point of this domain, system
(\ref{7}) has three stationary solutions (two asymptotically
stable and one unstable). For any point of the bifurcation curve,
except for the critical point $P_\textrm{c} \equiv
(q_\textrm{c}\,, \, \ell_\textrm{c})$, where $q_\textrm{c} = 1,\
\ell_\textrm{c} = \exp(-2) \approx 0.135$ is the critical
concentration, which is the common point of the branches and the
cusp of the second kind of the bifurcation curve, system (\ref{7})
has two stationary solutions (one is asymptotically stable and the
other one is two-fold). At the cusp, system (\ref{7}) has one
three-fold stationary solution. Curve M divides the domain of
bistability into two subdomains: for any point lying between
curves 1 and M, the system is stable in the first well and
metastable in the second one; for any point lying between curves 2
and M, the system is stable in the second well and metastable in
the first one. For any point of curve M, except for the point
$P_\textrm{c}\,$, both wells have equal depths and, hence, two
stable states of the system coexist and curve M is a curve of the
coexistence of two stable states of the system. Taking into
account thermal fluctuations in the system under study, we can
conclude that the motion in the plane of control parameters $(q,
\ell)$ along any line intersecting curve M is accompanied by the
transition of the system from one well into another (deeper) well
after the intersection of curve M. At the point of intersection of
the line and curve M,
the first-order phase transition occurs \cite{Gil}. %\looseness=-2

Given the explicit expression for the Maxwell concentration
(\ref{21}), we can easily plot the actual adsorption isotherms
based on the Maxwell principle (and called Maxwell adsorption
isotherms) on the basis of adsorption isotherms defined by
relations (\ref{16}), (\ref{17}). To this end, we first note that,
for $g > 4$, each stable branch of an adsorption isotherm has a
metastable part lying to the right (for the lower stable branch)
or to the left (for the upper stable branch) of the point of the
branch with abscissa $\ell = \ell_\textrm{M}\,$. A Maxwell
adsorption isotherm consists of the corresponding adsorption
isotherm defined by (\ref{16}), (\ref{17}) without the unstable
branch and the metastable parts of the stable branches and the
segment $\textrm{AB}$ of the vertical straight line $\ell =
\ell_\textrm{M}$ connecting the stable branches. Hence, a Maxwell
adsorption isotherm is an adsorption isotherm of a system of
adparticles having one stable state for any concentration $\ell$
except for the single concentration $\ell = \ell_\textrm{M}$ for
which the system has two energy-equivalent states and which,
following \cite{Do}, may be called the phase transition
concentration. Note that Maxwell adsorption isotherms thus
constructed are similar to the well-known adsorption isotherms
taking into account attractive lateral interactions between
adparticles on a nondeformable adsorbent (see, e.g.,
\cite{Zhd,AdC,Tov,RuE,Do}). Furthermore, the Maxwell concentration
(\ref{21}) agrees with the corresponding expression for the phase
transition pressure for the classical Fowler-Guggenheim
adsorption isotherms in the case of attractive lateral
interactions of adparticles on a nondeformable adsorbent \cite{Do}
if the maximum increment of the activation energy for desorption
of adparticles $|V^\textrm{a}|$ caused by the adsorbent
deformation in (\ref{8}) is replaced by the modulus of the energy
of attractive lateral interaction of two neighboring adparticles
for $\theta = 1$ multiplied
by the coordination number of the lattice of adsorption sites. %\looseness=-2

The adsorption isotherm and the Maxwell adsorption isotherm
consisting of stable branches of the adsorption isotherm without
metastable parts and the segment $\textrm{AB}$ of the vertical
straight line $\ell = \ell_\textrm{M}$ connecting the branches in
figure~\ref{pfig5} clearly illustrate the essential difference
(the presence and absence of a hysteresis loop, respectively) in
the behavior of the amount of adsorbed substance with $\ell$. The
vertical part (the segment $\textrm{AB}$) of the Maxwell
adsorption isotherm indicates the coexistence of two stable states
of the system under study and corresponds to the first-order phase
transition in the system. It is easy to see that the areas of the
domains enclosed by the adsorption isotherm and the segment
$\textrm{AB}$ to the left and to the right of the segment
$\textrm{AB}$ are different. Nevertheless, the equality of these
areas is proved. Furthermore, for any $g > 4$ (and, hence, the
well-known Maxwell rule of equal areas is true for the adsorption
isotherms on a deformable adsorbent) if $\ln \ell $ is laid off
along the abscissa axis instead of $\ell$ (see also the Maxwell
rule of equal areas for the adsorption isotherms taking into
account attractive lateral interactions between adparticles on a
nondeformable adsorbent \cite{RuE}). It is also worth noting that
the ordinate of the point of intersection of the unstable branch
of the adsorption isotherm and the vertical straight line $\ell =
\ell_\textrm{M}$ is equal to $1/2$ for any value of $g$ due to the
evenness of the Maxwell
potential $V^\textrm{M}(\xi)$ about $\xi = 1/2$. %\looseness=-2

According to the remark before figure~\ref{pfig1}, the thick
curve in figure~\ref{pfig5} also describes the actual (Maxwell)
behavior of the adsorption-induced surface
normal relaxation of a solid adsorbent with concentration $\ell$. %\looseness=-2

%\newpage

%%---------------------------------- Section 3 ------------------------------

\section{Probability density}  \label{Stationary}

To investigate transitions of the bistable system between its
stable states with variation in the control parameters  $\ell$ and
$g$, we take into account thermal fluctuations in the system
introducing a Langevin force~$\mathcal{F}(t)$ in the right-hand
side of the deterministic equation of motion of oscillator
(\ref{13}), \cite{Ryt1,Gar,Ris} which yields the stochastic
differential equation
\begin{equation}
 \alpha \frac{\rd  x}{\rd t} = F(x) + \mathcal{F}(t)\,.
 \label{28}
\end{equation}
\noindent  The random force $\mathcal{F}(t)$ has the properties of a white
noise: %\looseness=-1
\begin{equation}
 \left\langle\mathcal{F}(t) \right\rangle = 0\,, \qquad
 \left\langle\mathcal{F}(t)\, \mathcal{F}(t') \right\rangle = 2k_\textrm{B} T\alpha \,\delta(t-t'),
 \label{29}
\end{equation}
\noindent  where the angular brackets  $\,\left\langle \ldots
\right\rangle\,$ denote the averaging over an ensemble of
realizations of the random force $\mathcal{F}(t)$, the quantity
$2k_\textrm{B} T\alpha$ in the correlation function in (\ref{29})
is the intensity of the Langevin force, and $\delta(x)$ is the
Dirac $\delta$-function.

Following \cite{Gar}, we denote random variables by capital
letters and their values by small letters (for example, $X(t)$ is
a realization of the dynamical variable $x$ at the time  $t$).
Equation~(\ref{28}) can be reduced to the Fokker-Planck equation
for the probability density $p(x,t) = \left\langle\delta\left(x -X(t)\right)
\right\rangle$ of the coordinate of oscillator
\cite{Ryt1,Gar,Ris}, which also describes the adsorption-induced
surface normal relaxation of a solid adsorbent with regard for
thermal
fluctuations, %\looseness=-1
\begin{equation}
\alpha\,\frac{\partial p(x,t)}{\partial t} = \frac{\partial
}{\partial x} \, \Biggl [\frac{\rd  U(x)}{\rd x}\, p(x,t) +
k_\textrm{B} T \, \frac{\partial p(x,t)}{\partial x} \Biggr ]\,.
\label{30}
\end{equation}

Given the quantity $p(x,t)$ as a solution of the Fokker-Planck
equation (\ref{30}), the probability density~$p(\xi,t)$ of the
random variable $\,\Xi\,$ is expressed in terms of $p(x,t)$ as
follows:
\begin{equation}
p(\xi,t) = |x_{\textrm{max}}|\, p(x,t)\big|_{\,x = x_{\textrm{max}}\, \xi}\,.
\end{equation}

By virtue of (\ref{9}), the joint probability density $p(\xi,\theta; t)$ of the random variables and  $\Xi$ and $\Theta$
has the form
\begin{equation}
p(\xi,\theta; t) = p(\xi,t)\,\delta\big(\theta - f(\xi)\big)\, ,  \qquad
  \theta \in [0,1],
  \label{32}
\end{equation}
where the $\delta$-function on the right-hand side of (\ref{32})
is the conditional probability density $p(\theta|\,\xi; t) =
\delta(\theta - f(\xi))$ with the sharp value for $\theta =
f(\xi)$ and $f(\xi)$ is the deterministic function equal to the
right-hand side of relation (\ref{9}).

We first consider the stationary case. Under the natural boundary conditions,
the stationary probability density $p(x)$ has the Boltzmann distribution
\cite{Gar,Ris}
\begin{equation}
p(x) = \mathcal{N}\, \exp\left[-\frac{U(x)}{k_\textrm{B}
T}\right], \label{33}
\end{equation}
where $\mathcal{N}$ is the normalization constant defined as follows:
\begin{equation}
\mathcal{N}^{-1}\, = \int\limits_{-\infty}^\infty \!\! \rd x\,
   \exp\left[-\frac{U(x)}{k_\textrm{B} T}\right].
\end{equation}

In view of (\ref{33}), (\ref{12}), and (\ref{15}), we get
\begin{equation}
p(\xi) = \widetilde{\mathcal{N}}\, \exp\left[-\frac{g}{2}\,
V(\xi)\right], \label{35}
\end{equation}
\noindent  where
\begin{equation}
\widetilde{\mathcal{N}} = |x_{\textrm{max}}|\, \mathcal{N} =
   \sqrt{\frac{g}{2\pi}}\,\frac{1 + \ell}{1 + \lambda}\,, \qquad
   \lambda = \frac{\ell}{\ell_\textrm{M}}\,.
\end{equation}

According to (\ref{35}), the functions $p(\xi)$ and $V(\xi)$ have
extrema at the same points, moreover, if  $V(\xi)$ has a minimum
(maximum) at some point, then $p(\xi)$ has a maximum (minimum) at
this point \cite{Gar,Ris}. Hence, the stationary probability
density $p(\xi)$ is single-modal for $g \leqslant 4$ and for $g >
4$, $\ell \notin [\ell^\textrm{b}_1\,, \, \ell^\textrm{b}_2]$ and
bimodal for $g > 4$, $\ell \in (\ell^\textrm{b}_1\,, \,
\ell^\textrm{b}_2)$.
%\looseness=-1

By using the explicit expression (\ref{12}) for $V(\xi)$, we
obtain
\begin{equation}
p(\xi) = \sqrt{\frac{g}{2\pi}}\, \exp{\left(-\frac{g\,\xi^2}{2}\right )}\,
 \frac{1 + \ell\, \exp{(g\,\xi)}}{1 + \lambda}\,, \qquad
p(x) = \frac{1 + \ell\, \exp{(b\,x)}}{1 + \lambda}\, p_0(x),
\label{37}
\end{equation}
\noindent  where
\begin{equation}
p_0(x) = \frac{1}{\sqrt{2\, \pi}\, \sigma_0}\, \exp{\left(-\frac{x^2}{2\,
\sigma_0^2}\right )}
\end{equation}
\noindent is the Gaussian distribution of the probability density for a linear
oscillator with zero mean ($\left\langle X \right\rangle\, = 0$) and the
variance $\sigma_0^2 = k_\textrm{B} T/\varkappa$.

We first consider the single-modal stationary probability density $p(x)$. In
this case, the random variable $X$ has the nonzero mean
\begin{equation}
\left\langle X \right\rangle\, =\, \frac{\lambda}{1 + \lambda}\:
x_{\textrm{max}}\,, \label{39}
\end{equation}
\noindent which yields $\,\sign \left\langle X \right\rangle\, = \sign\,\chi$
and, hence, the maximum of the probability density $p(x)$ shifts in the
direction of the action of the adsorption-induced force; the variance $\sigma^2
\equiv\, \left\langle(X\, - \langle X \right\rangle)^2\rangle$,
%\looseness=-1
\begin{equation}
\sigma^2 = \left[1 + g\,\frac{\lambda}{(1 + \lambda)^2}\right]\,
\sigma_0^2\,, \label{40}
\end{equation}
\noindent  which is greater than $\sigma_0^2$ for any values of the
concentration and the coupling parameter and, for a fixed value of $g$, reaches
its maximum value equal to $(1 + q)\,\sigma_0^2\,$ for $\lambda = 1$; the
asymmetry ratio $S = \langle(X\, - \left\langle X
\right\rangle)^3\rangle/\sigma^3$,  %\looseness=-1
\begin{equation}
S = g^{3/2}\, \frac{\lambda\,(1 - \lambda^2)(1 + \lambda)^2}
    {[(1 + \lambda)^2 + g\, \lambda]^3}\,\sign\,\chi,
\end{equation}
\noindent  and, hence, $\,\sign\,S = \sign\,\big((\ell_\textrm{M} -
\ell)\,\chi\big)$, which implies the change in the sign of the
asymmetry ratio in crossing the concentration $\ell_\textrm{M}\,$;
and the excess (flatness)  $E =\langle(X\, -
\left\langle X \right\rangle)^4\rangle/\sigma^4 - 3$, %\looseness=-1
\begin{equation}
E = g^2\, \frac{\lambda\,(1 + \lambda^2 - 4\, \lambda)}{[(1 +
\lambda)^2 + g\, \lambda]^2}\,. \label{42}
\end{equation}
\noindent  According to (\ref{42}), the probability density $p(x)$
is flat-topped ($E<0)$ for $\ell \in (\ell_-\,, \, \ell_+)$,
where $\ell_\pm  = (2 \pm \sqrt{3})\, \ell_\textrm{M}\,$, or
peaked ($E>0$) for $\ell \notin [\ell_-\,, \, \ell_+]$ relative to the
Gaussian distribution with mean (\ref{39}) and variance
(\ref{40}). For $\ell = \ell_-$ and $\ell = \ell_+\,$, $E= 0$ as
for this Gaussian distribution.
%\looseness=-1

In the Maxwell case ($\ell = \ell_\textrm{M}$), the probability
density $p(x)$ in (\ref{37})
denoted by $p_\textrm{M}(x)$ is simplified to the form %\looseness=-1
\begin{equation}
p_\textrm{M}(x) = \frac{\ell_\textrm{M}^{1/4}}{\sqrt{2\, \pi}\, \sigma_0}\,
 \exp{\left(-\frac{\tilde{x}^2}{2\, \sigma_0^2}\,\right)}\,
 \cosh{\left(\frac{b\,\tilde{x}}{2}\right)}\,, \qquad
 \tilde{x} = x -\frac{x_{\textrm{max}}}{2}\,,
\end{equation}
\noindent  which is an even function about $x = x_{\textrm{max}}/2$.

In the single-modal case ($g \leqslant 4$), we have $\left\langle
X \right\rangle = x_{\textrm{max}}/2$, $\sigma^2 = (1+q)
\sigma_0^2$, $S = 0$, and $E = -2q^2/(1+q)^2 < 0$. Hence,
$p_\textrm{M}(x)$ is a flat-topped distribution symmetric about
its maximum value equal to \, $\ell_\textrm{M}^{1/4}/(\sqrt{2\,
\pi}\, \sigma_0)$\, at $x = x_{\textrm{max}}/2$. %\looseness=-1

We now investigate the transition of the bistable system from one
stable state to another due to thermal fluctuations. This
transition occurs for the double-well potential $V(\xi)$ studied
above: the left-hand and right-hand wells with minimal values
$V_\textrm{L} \equiv V_1$ and $V_\textrm{R} \equiv V_3\,$,
respectively, at $\xi = \xi_1 \equiv \xi_\textrm{L}\,$ and $\xi =
\xi_3 \equiv \xi_\textrm{R}$ are separated by the barrier with
maximum value $V_\textrm{B} \equiv V_2$ at $\xi = \xi_2 \equiv
\xi_\textrm{B}\,$. By using the Gardiner representation \cite{Gar}
of the Kramers method \cite{Kra}, the following system of
equations is derived from the Fokker-Planck equation (\ref{30}):
%\looseness=-1
\begin{eqnarray}
 && \left\{
\begin{array}{l}
  \dfrac{\rd P_\textrm{L}(t)}{\rd t} = -k_\textrm{L}\, P_\textrm{L}(t)
     + k_\textrm{R}\, P_\textrm{R}(t),
      \vspace{3mm}  \\
  \dfrac{\rd P_\textrm{R}(t)}{\rd t} = -k_\textrm{R}\, P_\textrm{R}(t)
     + k_\textrm{L}\, P_\textrm{L}(t).
\end{array}\right.
\label{44}
\end{eqnarray}
\noindent  This system describes the evolution of the probabilities of the
presence of an oscillator to the left (in the left-hand well), $P_\textrm{L}(t)$,
and to the right (in the right-hand well), $P_\textrm{R}(t)$, of the point
$\xi_\textrm{B}$ at time $t$ under the assumption that the relaxation times of the
oscillator in the wells are much less than the mean transition times between
the wells %\looseness=-2
\begin{equation}
 P_\textrm{L}(t) = \int\limits_{-\infty}^{\xi_\textrm{B}} \!\! \rd\xi\, p(\xi, t),  \qquad
 P_\textrm{R}(t) = \int\limits_{\xi_\textrm{B}}^\infty \!\! \rd\xi\, p(\xi, t),   \qquad
 P_\textrm{L}(t) + P_\textrm{R}(t) = 1.
\end{equation}
\noindent  Here,
\begin{equation}
 k_i = \frac{1}{g\mu\, P_i\, \tau_{\text{r}}}\,,  \qquad i = \textrm{L,\,R},
 \label{46}
\end{equation}
\noindent is the coefficient of the escape rate of an oscillator
from the $i$th well ($i = \textrm{L,\,R}$) derived under the
assumption of a vanishing probability of the presence of the
oscillator in the well outside a small neighborhood of $\xi_i\,$,
$i = \textrm{L,\,R}$, and $P_i$ is the probability of the presence
of the oscillator in the $i$th well ($i = \textrm{L,\,R}$) in the
stationary case, i.e.,
\begin{equation}
 P_\textrm{L} = \int\limits_{-\infty}^{\xi_\textrm{B}} \!\! \rd\xi\, p(\xi)\,,  \qquad
 P_\textrm{R} = \int\limits_{\xi_\textrm{B}}^\infty \!\! \rd\xi\, p(\xi)\,
\end{equation}
\noindent  and
\begin{equation}
 \mu = \int\limits_{\xi_\textrm{L}}^{\xi_\textrm{R}} \!\! \frac{\rd \xi}{p(\xi)}\,.
 \label{48}
\end{equation}
The solution of system (\ref{44}) has the form
\begin{equation}
 P_i(t) = P_i^0\, \exp{\left(-\frac{t}{T_+}\right)} + P_i\,
  \left[1 - \exp{\left(-\frac{t}{T_+}\right)}\right]\,,
  \qquad i = \textrm{L,\,R},
\end{equation}
\noindent where $P_i^0$ is the initial value of the probability
$P_i(t), \, i = \textrm{L,\,R}$, for $t = 0$,  $k_+ = k_\textrm{L}
+ k_\textrm{R}\,$, and
\begin{equation}
T_+ = 1/k_+ = g\mu\, P_\textrm{L}\, P_\textrm{R}\, \tau_{\text{r}}
\end{equation}
\noindent  is the relaxation time of the quantity $P_i(t)$, $i =
\textrm{L,\,R}$.

According to (\ref{44}), the mean transition time from one stable
state at $\xi = \xi_i$ to another at $\xi = \xi_j\,$, where $i,j =
\textrm{L,\,R}$, $i\neq j$, is defined as $T_{i\, \rightarrow\, j}
= 1/k_i$ and, with regard for (\ref{46}), has the
form %\looseness=-1
\begin{equation}
 T_{i\, \rightarrow\, j} = g\mu\, P_i\, \tau_{\text{r}}\,,  \qquad i,j = \textrm{L,\,R}, \qquad i\neq j,
 \label{51}
\end{equation}
which yields the well-known relationship \cite{Kam} between the
probabilities $P_i$ and the mean transition times~$T_{i\rightarrow j}$:
\begin{equation}
 \frac{P_\textrm{L}}{P_\textrm{R}} =
 \frac{T_{\textrm{L}\, \rightarrow\, \textrm{R}}}{T_{\textrm{R}\, \rightarrow\, \textrm{L}}}\,.
\end{equation}
In the parabolic approximation \cite{Gar,Kam} of the potential
$U(\xi)$ (\ref{12}) in the neighborhoods of its extrema, the mean
transition times $T_{i\, \rightarrow\, j}$ are estimated as
follows:
\begin{equation}
 T_{i\, \rightarrow\, j} \approx \tau_{\text{r}}\, \frac{2\pi}{\sqrt{L(\xi_i,g)\, |L(\xi_2,g)|}}\,
 \exp{\Bigl(\frac{g}{2}\, \Delta_{\,\textrm{B},\,i}\Bigr)},   \qquad i,j = \textrm{L,\,R}, \qquad  i\neq j,
 \label{53}
\end{equation}
\noindent where
\begin{equation}
 \Delta_{\,i,j} = V_i - V_j = (\xi_i - \xi_j)\,(\xi_i + \xi_j -2)\,
 + \frac{2}{g}\, \ln{\frac{\xi_i}{\xi_j}}\,, \qquad  i,j = \textrm{L,\,R,\,B},
 \label{54}
\end{equation}
\begin{equation}
L(\xi,g) = 1 + g\xi\,(\xi -1)
\end{equation}
\noindent  is an even function about  $\xi = 1/2$ such that, for
$g > 4$, $1 >
L(\xi_i\,,g) > 0$, $i = \textrm{L,\,R},$ and $L(\xi_{\text{B}}\,,g) < 0$.  %\looseness=-1

In this approximation, the quantity
\begin{equation}
\tau_i = \frac{\tau_{\text{r}}}{L(\xi_i,g)}\,, \qquad  i =
\textrm{L,\,R}, \label{56}
\end{equation}
has the sense of the mean relaxation time of an overdamped
oscillator in the left-hand ($i = \textrm{L}$) or in the
right-hand ($i = \textrm{R}$) parabolic potential well centered at
$\xi = \xi_i$ and derived from (\ref{12}). Hence, in this
approximation, the restoring force constant $\varkappa$ is simply
replaced by $\varkappa_i$ = $\varkappa\, L(\xi_i\,,g)$, $i =
\textrm{L,\,R}$, that already depends on $\ell$
and $g$, which yields $\tau_i > \tau_{\text{r}}\,$, $i = \textrm{L,\,R}$. %\looseness=-1

If the concentration $\ell \in (\ell^\textrm{b}_1\,, \,
\ell^\textrm{b}_2)$ approaches the end point of the interval of
bistability ($\ell^\textrm{b}_1$ or $\ell^\textrm{b}_2$), then
$L(\xi_\textrm{R}\,,g)$ or $L(\xi_\textrm{L}\,,g)$, respectively,
tends to zero. According to (\ref{56}), this is accompanied by an
essential increase in the relaxation time $\tau_\textrm{R}$ (in
the first case) or $\tau_\textrm{L}$ (in the second case) due to
the flattening of the corresponding well. This indicates that the
parabolic approximation is false in small neighborhoods of the
bifurcation concentrations $\ell^\textrm{b}_1$ and
$\ell^\textrm{b}_2$ because the very shallow right-hand (in the
first case) and left-hand
(in the second case) wells are almost flat-bottomed. %\looseness=-1

Formally introducing the quantity
\begin{equation}
\tau_\textrm{B} =
\frac{\tau_{\text{r}}}{|\textrm{L}(\xi_\textrm{B}\,,g)|}\,,
\end{equation}
which can be regarded as the mean relaxation time of an overdamped
oscillator in the parabolic potential well centered at $\xi =
\xi_\textrm{B}$ and derived from (\ref{12}) with
$|L(\xi_\textrm{B}\,,g)|$ instead of $L(\xi_\textrm{B}\,,g)$, we
can express the mean transition times $T_{i\, \rightarrow\, j}$
(\ref{53}) in terms of the relaxation times of an overdamped
oscillator in the corresponding
parabolic potential wells as follows: %\looseness=-1
\begin{equation}
 T_{i\, \rightarrow\, j} \approx  2\pi\, \sqrt{\tau_i\, \tau_\textrm{B}}\,
 \exp{\Bigl(\frac{g}{2}\, \Delta_{\,\textrm{B}\,,\,i}\Bigr)}\,,
   \qquad i,j = \textrm{L,\,R}, \qquad  i\neq j,
   \label{58}
\end{equation}
which agrees with the classical Kramers formula in the overdamped case
\cite{Kra}.

The assumption of two time scales (the short time scale for the
relaxation of the oscillator in the well where it was at the
initial time and at the long-time scale for the transition of the
oscillator from one stable state to another across the unstable
state) used in the derivation of system (\ref{44}) imposes the
following condition on the Arrhenius factor: %\looseness=-1
\begin{equation}
 \exp{\Bigl(\frac{g}{2}\, \Delta_{\,\textrm{B},\,i}\Bigr)}  \gg  \frac{1}{2\pi}\,
 \sqrt{\frac{\tau_i}{\tau_\textrm{B}}}\,,   \qquad i = \textrm{L,\,R}\,.
\end{equation}

Substituting (\ref{54}) in (\ref{53}) and (\ref{58}), we express
the ratio of the mean transition times between the wells in terms
of the coordinates of their minima
$\xi_\textrm{L}\,$ and $\xi_\textrm{R}\,$ as follows: %\looseness=-1
\begin{eqnarray}
 \frac{T_{\textrm{L}\, \rightarrow\, \textrm{R}}}
  {T_{\textrm{R}\, \rightarrow\, \textrm{L}}}
  = \sqrt{\frac{\tau_\textrm{L}}{\tau_\textrm{R}}}\,
 \exp{\Bigl(\frac{g}{2}\, \Delta_{\,\textrm{R,\,L}}\Bigr)}=
  \sqrt{\frac{\tau_\textrm{L}}{\tau_\textrm{R}}}\, \frac{\xi_\textrm{R}}{\xi_\textrm{L}}\,
  \exp{\Bigl[\frac{g}{2}\, (\xi_\textrm{R}-\xi_\textrm{L})\,
  (\xi_\textrm{R}+\xi_\textrm{L}-2) \Bigr]}\,.
\end{eqnarray}

For the Maxwell potential (\ref{22}), we get $\xi_\textrm{B} =
1/2$, the quantity $L(\xi_\textrm{B}\,,g) = 1 - q$ \  is
independent of the concentration and, hence, $\tau_\textrm{B} =
\tau_{\text{r}}/(q-1)$,
\begin{equation}
 L(\xi_\textrm{L}\,,g) = L(\xi_\textrm{R}\,,g) \equiv L_\textrm{M}(g) =
  1-q\left(1- \frac{\eta^2}{q^2} \right)\,, \qquad
\tau_\textrm{L} = \tau_\textrm{R} \equiv \tau_\textrm{M} =
  \frac{\tau_{\text{r}}}{\textrm{L}_\textrm{M}(g)}\,,
  \label{61}
\end{equation}
\noindent where, as above, $\eta$ is a positive solution of
equation~(\ref{24}), equal mean transition times between the
wells,  $T_{\textrm{L}\, \rightarrow\, \textrm{R}} =
T_{\textrm{R}\, \rightarrow\, \textrm{L}} \equiv T_\textrm{M}$
and, with regard for (\ref{54}), (\ref{58}), and (\ref{61}),
\begin{equation}
 T_\textrm{M}  \approx  2\pi\,\sqrt{\tau_\textrm{M}\, \tau_\textrm{B}}\, \cosh{\eta} \,
      \exp{\left(-\frac{\eta^2}{2q}\right)}\,.
      \label{62}
\end{equation}

It is worth noting that expression (\ref{62}) cannot be used if
$q$ is close to 1 because, in this case, the assumptions of
derivation of $T_\textrm{M}$ are violated.

For large values of the coupling parameter, $q \gg 1$, the
quantity $\eta \approx q\, (1 - 2\, \ell_\textrm{M})$, which
yields $\tau_\textrm{M} \approx \tau_{\text{r}}\,$,
$\tau_\textrm{B} \approx \tau_{\text{r}}/q$,
\begin{equation}
 T_\textrm{M} \approx \tau_{\text{r}}\, \frac{\pi}{\sqrt{q}}\, \exp{\Bigl(\frac{q}{2}\Bigr)}\,.
 \label{63}
\end{equation}
\noindent  In terms of the potential $U(x)$, we have $U(x=
x_{\textrm{max}}/2)/k_\textrm{B} T \approx q/2$ \ for  $q \gg 1$,
which leads to the exponential growth of the mean transition time
(\ref{63}) with an increase in the coupling parameter $g$.
%\looseness=-2

Relations for the mean transition times derived in the general case are given
in Appendix. %\looseness=-1

%\newpage

%%-------------------------------- Conclusions ------------------------------

\section{Conclusions}  \label{Conclusions}

In the present paper, we have investigated the problem of adsorption of a gas
on the flat surface of a solid deformable nonporous adsorbent with regard for
thermal fluctuations and analyzed the effect of thermal fluctuations on the
normal displacement of the adsorbent surface and, hence, on the amount of
adsorbed substance. We have derived explicit expressions for the mean
transitions times between stable states of the bimodal system and investigated
the dependence of these times on the values of the coupling parameter and the
gas concentration. %\looseness=-1

According to the established results, the behavior of the system under study is
most interesting for the case where the system is bistable. For the bistability
of a system, the coupling parameter must exceed the threshold value. Thus,
first of all, the value of the coupling parameter for the investigated
adsorbent--adsorbate system should be calculated. However, this requires an
additional information because the coupling parameter is expressed in terms of
the phenomenological constant adsorption-induced force $\chi$. Nevertheless,
this unknown force (and, hence, the coupling parameter) can be expressed in
terms of the maximum change in the first interplanar spacing $x_{\textrm{max}}$ in the
case of the total monolayer coverage of the adsorbent surface by adparticles:
%\looseness=-2
 \begin{equation}
 \chi = \varkappa\, x_{\textrm{max}}\,,  \qquad
 g = \varkappa\, x^2_{\textrm{max}}/k_\textrm{B} T.
 \label{64}
\end{equation}
Having the experimental value of $x_{\textrm{max}}$  for such an
adsorbent--adsorbate system measured with proper accuracy, the
value of the coupling parameter is easily calculated by using the
second relation in (\ref{64}). Given a base of experimental data
(lacking at present) for $x_{\textrm{max}}$  for various
adsorbent--adsorbate systems, one can select systems for which the
established effects of bistability of the system caused by the
adsorption-induced
deformation of the adsorbent are possible.  %\looseness=-1

Since the coupling parameter is proportion to $x^2_{\textrm{max}}$
(\ref{64}), experiments with solid nonporous adsorbents with flat
surface should be performed for materials with a considerable
value of the adsorption-induced normal
displacement of the surface. %\looseness=-1

Note that the results established in the present paper have been obtained for
the model imposing the restrictions on the values of characteristic times (the
average time between collisions of gas particles with an adsorption site, the
average residence time of an adparticle on the surface, and the relaxation time
of a bound adsorption site) and the friction coefficient. %\looseness=-1

Since the proposed model of adsorption on a deformable solid adsorbent does not
take into account lateral interactions between adparticles, it is of interest
to generalize this model to the case taking into account the joint action of
both factors (adsorption-induced deformation of the adsorbent and lateral
interactions between adparticles).  %\looseness=-2

%\newpage

%%------------------------------- Acknowledgments ------------------------------

\section*{Acknowledgements}

The author expresses deep gratitude to Prof. Yu.\,B.~Gaididei for
valuable remarks and useful discussions of results.

%\newpage

%%---------------------------------- Appendix -------------------------------

%\numberwithin{equation}{section}   % Enumeration in sections:  (1.1), (1.2) in Sec.~1;

\section*{Appendix}

The mean transition times $T_{i\, \rightarrow\, j}$ (\ref{58})
derived above in the parabolic approximation can also be
determined for the general case. By using expression (\ref{12})
for  $V(\xi)$ and relations (\ref{35})--(\ref{37}),
(\ref{46})--(\ref{48}), and (\ref{51}), we obtain the following
relations for stationary probabilities:
%\looseness=-2
\begin{equation}
 P_\textrm{R} = \frac{1}{2\,(1 +\lambda)}\,
 \Bigl[\erfc\left(\sqrt{2\,q}\,\xi_\textrm{B}\right) + \lambda\, \erfc\left(\sqrt{2\,q}\,(\xi_\textrm{B} -1)\right) \Bigr ]\,,
 \qquad  P_\textrm{L} = 1 - P_\textrm{R}\,,
 \tag{A.1}
 \label{A.1}
\end{equation}
\noindent and the mean transition times
\begin{align}
 T_{i\, \rightarrow\, j} &= \tau_{\text{r}}\, P_i\, 2(1+\lambda)\, \sqrt{\pi}
  \sum_{n = 0}^{\infty} \frac{(-1)^n}{\ell^{n+1}}\,
 \exp{\left\{-2q\,\xi_\textrm{L}\,[2(n+1)-\xi_\textrm{L}]\right\}}\Bigl\{F\left(\sqrt{2\,q}\,(n+1- \xi_\textrm{L})\right)
   \nonumber  \\
  &\quad -\exp{\left\{2q\,(\xi_\textrm{L}-\xi_\textrm{R})\,[2(n+1)-\xi_\textrm{L}-\xi_\textrm{R}]\right\}}\,
     F\left(\sqrt{2\,q}\,(n+1 - \xi_\textrm{R})\right) \Bigr\}\,,   \qquad i,j = \textrm{L,\,R}, \qquad   i\neq j\,.
   \tag{A.2}
   \label{A.2}
   \end{align}
\noindent  Here,
\begin{equation*}
 \erfc(z) = \frac{2}{\sqrt{\pi}}\, \int\limits_{z}^{\infty} \!\! \rd t\, \exp{\left(-t^2\right)}
\end{equation*}
\noindent is the complementary error function \cite{AbS}, the special function
\cite{Leb}
\begin{equation*}
 F(z) = \exp{(-z^2)}\, \int\limits_{0}^{z} \!\! \rd t\, \exp{\left(t^2\right)},
\end{equation*}
\noindent which is expressed in terms of the error function of imaginary
argument, is bounded for all real $z$, has the maximum value $\approx 0.541$
for $z \approx 0.924$, the known power expansion for $|z| < \infty$, and the
asymptotic formula $F(z) \approx 1/(2z)$ as  $z \rightarrow \infty$.
%\looseness=-1

In the Maxwell case ($\ell = \ell_\textrm{M}$), relations
(\ref{A.1}) yield equal values of the probabilities of the
presence of the oscillator in both wells: $P_\textrm{L} =
P_\textrm{R} = 1/2$, which is quite natural for the symmetric
Maxwell double-well potential (\ref{22}). Relations (\ref{A.2})
are reduced to the Maxwell
mean transition time  %\looseness=-1
\begin{equation}
 T_\textrm{M}  = \tau_{\text{r}}\,  4\sqrt{\pi}\, \exp{\Bigl(\frac{q}{2}\Bigr)}\, \sum_{n = 0}^{\infty}
  (-1)^n\, S_n(q),
  \tag{A.3}
  \label{A.3}
\end{equation}
\noindent  where
\begin{equation}
 S_n(q) = F(\beta_n) - \exp{[-\gamma\,(2\,\beta_n - \gamma)]}\,
          F(\beta_n -\gamma)\,,  \qquad
 \beta_n = \sqrt{\frac{q}{2}}\,(2n+1),  \qquad
 \gamma = \frac{\eta}{\sqrt{2\,q}}\,,
 \tag{A.4}
 \label{A.4}
\end{equation}
\noindent  $\eta$  is a positive solution of equation~(\ref{24}).

For $q \gg 1$, relation (\ref{A.3}) is simplified to the form
\begin{equation}
 T_\textrm{M} \approx \tau_{\text{r}}\, \frac{\pi^{3/2}}{\sqrt{2\,q}}\, \exp{\Bigl(\frac{q}{2}\Bigr)}\,.
 \tag{A.5}
 \label{A.5}
\end{equation}

Comparing (\ref{A.5}) and (\ref{63}) derived for the general case
and in the parabolic approximation, respectively, we see that, for
large values of $q,$ the mean
transition times differ only by the factor $\sqrt{\pi/2}$\,. %\looseness=-2

%\newpage

%%------------------------------- References ------------------------------

{\small \topsep 0.6ex

}

%\subsection*{Ukrainian part}
%\label{ua-part}
\subsection*{}

\ukrainianpart

\title{Індукована адсорбцією нормальна релаксація поверхні твердого адсорбенту}
\author{О.С.~Усенко}
\address{Інститут теоретичної фізики ім.~М.М.~Боголюбова НАН України, \\
 вул.~Метрологічна 14~б, 03680 Київ, Україна}

\makeukrtitle

\begin{abstract}
 \tolerance=3000%
Досліджено адсорбцію газу на плоскій поверхні твердого адсорбенту, який
деформується при адсорбції, враховуючи теплові флуктуації, і детально
проаналізовано вплив теплових флуктуацій на деформацію адсорбенту. Отримано
умову співіснування двох станів бістабільної системи адсорбованих частинок.
Встановлено особливості індукованою адсорбцією нормальної релаксації поверхні
адсорбенту, які обумовлені тепловими флуктуаціями. Знайдено середні часи
переходів між двома стійкими станами бістабільної системи в параболічному
наближені і в загальному випадку.
\keywords  адсорбція, ізотерма, деформація, флуктуації, бістабільність,
гістерезис

\end{abstract}

\end{document}